\pdfoutput = 1

\documentclass[12pt,a4paper]{article}

\usepackage{amsmath}
\usepackage{amssymb}
\usepackage{cite}
\usepackage{hyperref}
\usepackage{graphicx}
\usepackage{booktabs}
\usepackage[caption=false]{subfig}
\usepackage{comment}
\newcommand{\lsim}{\mathrel{\hbox{\rlap{\hbox{\lower4pt\hbox{$\sim$}}}\hbox{$<$}}}}

\usepackage{xspace}

\newcommand{\Bbar}{\kern 0.18em\overline{\kern -0.18em B}{}\xspace}

\newcommand{\Kbar}{\kern 0.18em\overline{\kern -0.18em K}{}\xspace}

\usepackage{xcolor}
\newcommand{\kkv}[1]{{\color{blue}#1}}

\begin{document}


\begin{titlepage}

\vspace*{-0.0truecm}

\begin{flushright}
Nikhef-2022-020\\
\today
\end{flushright}

\vspace*{0.3truecm}

\begin{center}
{\Large \bf \boldmath Zooming into CP violation in $B_{(s)}\to hh$ Decays}
\end{center}

\vspace{0.9truecm}

\begin{center}
{\bf Robert Fleischer,\,${}^{a,b}$  Ruben Jaarsma,\,${}^{a}$ and  K. Keri Vos\,${}^{a,c}$}

\vspace{0.5truecm}

${}^a${\sl Nikhef, Science Park 105, NL-1098 XG Amsterdam, Netherlands}

${}^b${\sl  Department of Physics and Astronomy, Vrije Universiteit Amsterdam,\\
NL-1081 HV Amsterdam, Netherlands}

{\it $^c$Gravitational 
Waves and Fundamental Physics (GWFP),\\ 
Maastricht University, Duboisdomein 30,\\ 
NL-6229 GT Maastricht, the
Netherlands}\\[0.3cm]

\end{center}

\vspace*{1.7cm}


\begin{abstract}
\noindent
The LHCb collaboration has recently reported the first observation of  CP violation in the penguin-dominated $B^0_s\to K^-K^+$ decay and further new measurements, indicating differences
between the direct CP asymmetries of both the $B^0_s\to K^-K^+$, $B^0_d\to\pi^-K^+$ and the $B^0_d\to \pi^-\pi^+$, $B^0_s\to K^-\pi^+$ modes. We show that these puzzling differences can be accommodated through sizeable penguin annihilation and exchange topologies in the Standard Model, and constrain them. Utilising the $U$-spin symmetry, we extract the angle $\gamma$ of the unitarity triangle from the CP asymmetries in the $B^0_s\to K^-K^+$, $B^0_d\to \pi^-\pi^+$ system alone, finding $\gamma=(65^{+11}_{-7})^\circ$, in perfect agreement with the determination from tree-level $B\to DK$ decays. The $B^0_s$--$\bar B^0_s$ mixing phase $\phi_s$ can be extracted from CP violation measurements in $B^0_s\to K^-K^+$ in a clean way. We present a new strategy and extract $\phi_s=-(3.6\pm 5.4)^\circ$. This result is in agreement with the determination from $B^0_s\to J/\psi \phi$ decays. New CP-violating contributions would influence these determinations differently. Hence it is interesting to keep monitoring both as the experimental picture sharpens.   

\end{abstract}


\vspace*{2.1truecm}

\vfill

\noindent

\end{titlepage}


\thispagestyle{empty}

\vbox{}



\setcounter{page}{1}

\newpage
\section{Introduction}
Charmless two-body $B$ decays are powerful probes of CP violation (see e.g.~\cite{Gronau:1995hm,Neubert:1998pt,Fleischer:1999pa,Buras:1998rb,Fleischer:2017vrb,Fleischer:2018bld,Huber:2021cgk }). These $B\to h h$ modes, where $h = \pi, K$, haven been studied in a continuous effort both from the theoretical and experimental sides. Especially, the penguin-dominated $B_s^0\to K^-K^+$ decay plays a key role \cite{Fleischer:1999pa,Fleischer:2007hj, Fleischer:2010ib,Ciuchini:2012gd, Fleischer:2016ofb,Fleischer:2016jbf,Nir:2022bbh} because its CP asymmetries can be used to determine the angle $\gamma$ of the unitarity traingle (UT) and the $B_s^0$--$\bar{B}_s^0$ mixing phase $\phi_s$. 

The LHCb collaboration recently reported the first observation of CP violation in this decay, and updated the other modes \cite{LHCb:2020byh}. The precision of the measured CP asymmetries of these decays has become impressively high, 
thereby putting these decays once again into the spotlight. In this paper, we will present the first determinations of $\gamma$ and $\phi_s$ from the CP asymmetries in $B_s^0\to K^-K^+$. 

The $U$-spin symmetry links the $B_s^0 \to K^-K^+$ decay to the $B_d^0\to \pi^-\pi^+$ channel. Together, these modes allow us to determine the corresponding CKM angle $\gamma$ via the interference of tree-level topologies and quantum-loop effects through QCD penguins. For the first time, we are able to determine $\gamma$ from these decays using only the CP asymmetries. This new determination should be compared with the determination from the pure tree-level $B\to D K$ decays, which would be influenced by possible new-physics interactions in a different way.

Beyond that, we note that the new LHCb measurements  \cite{LHCb:2020byh} reveal an interesting pattern. We find for the difference between the direct CP asymmetries in $B_s^0 \to K^- K^+$ and $B_d^0 \to \pi^- K^+$ decays: 
\begin{align}
	\mathcal{A}_{\rm CP}^{\rm dir}(B_s^0 \to K^- K^+) - \mathcal{A}_{\rm CP}^{\rm dir}(B_d^0 \to \pi^- K^+) &= 0.089 \pm 0.031 \ , \label{eq:aCPdirDiff1} 
	\end{align}
which differs from zero by $2.9 \,\sigma$. 
A similar difference is found in the $B_d^0 \to \pi^- \pi^+$ and $B_s^0 \to K^- \pi^+$ decays:
\begin{align}	
	\mathcal{A}_{\rm CP}^{\rm dir}(B_d^0 \to \pi^- \pi^+) - \mathcal{A}_{\rm CP}^{\rm dir}(B_s^0 \to K^- \pi^+) &= -0.095 \pm 0.040. \label{eq:aCPdirDiff2}
\end{align}
These channels only differ by their respective spectator quarks and by exchange ($E$) and penguin annihilation ($PA$) topologies that only enter in the first. These topologies are expected to play a minor role, and are usually neglected. Within this approximation the direct CP asymmetries of these decays would be equal. In fact, the difference in the CP asymmetries in \eqref{eq:aCPdirDiff1} and \eqref{eq:aCPdirDiff2} is challenging to explain via new-physics (NP) effects precisely because the decays originate from the same quark-level transition and only differ by their spectator quarks. It is therefore interesting to find out if the difference in the CP asymmetries can be accommodated by reasonable Standard Model (SM) exchange and penguin-annihilation contributions. Since the $E$ and $PA$ topologies are highly non-factorizable, they cannot be reliably computed and have to be determined from data. Therefore, we devise a strategy to determine the size of the $E$ and $PA$ contributions from experimental data using $U$-spin symmetry. 

The mixing-induced CP asymmetry of the $B_s^0\to K^-K^+$ decay also allows us to determine the mixing phase $\phi_s$ \cite{Fleischer:2016ofb,Fleischer:2016jbf}, complementary to values extracted from $B_s^0 \to J/\psi \phi$ (see e.g. \cite{Faller:2008gt,DeBruyn:2014oga,Barel:2020jvf, Barel:2022wfr}). Using the semileptonic $B_s^0$ and $B_d^0$ differential rates, $U$-spin breaking corrections only enter in theoretically favourable ratios multiplying small parameters. As measurements of the $B_s^0\to K^- \ell^+ \nu_\ell$ differential rate are currently not available, we will present a new strategy in which also form factors of the $B_d^0$ and $B_s^0$ decays enter in the form of an $SU(3)$-breaking ratio. 

This paper is organized as follows: In Sec.~\ref{sec:gammaDetermination} we discuss the $B_d^0\to \pi^-\pi^+, B_s^0\to K^- K^+$ $U$-spin system and use the CP asymmetries in these decays to extract the CKM angle $\gamma$. We then continue with the $U$-spin related $B_d^0\to \pi^- K^+, B_s^0 \to K^- \pi^+$ system and discuss a new strategy to determine the exchange and penguin-annihilation contributions from the experimental data. In Sec.~\ref{sec:phisDetermination}, we discuss a new strategy to determine $\phi_s$ from the $B_s^0 \to K^-K^+$ decay in light of our new insights. Finally, we conclude in Sec.~\ref{sec:Conclusion}.

\begin{table}
	\centering
	\begin{tabular}{l | l | c | c | c | c }
		 & & \multicolumn{4}{c}{Topologies} \\
		 \hline
		Mode & Notation & $T$ & $P$ & $E$ & $PA$ \\
		\hline
		\hline
		$B_d^0 \to \pi^- \pi^+$ & $d, \theta$ & x & x & x & x \\
		$B_s^0 \to K^- K^+$ & $d', \theta'$ & x & x & x & x \\
		\hline
		$B_s^0 \to K^- \pi^+$ & $\tilde{d}, \tilde{\theta}$ & x & x & & \\
		$B_d^0 \to \pi^- K^+$ & $\tilde{d}', \tilde{\theta}'$ & x & x & & \\
				\hline
		$B_s^0 \to\pi^- \pi^+$ & $\hat{d}, \hat{\theta}$ & &  & x & x \\
		$B_d^0 \to K^- K^+$ & $\hat{d}', \hat{\theta}'$ &  &  & x & x
	\end{tabular}
	\caption{Decay topologies contributing to the different $B \to h h$ modes ($h=\pi, K$) and the notation of the corresponding hadronic parameters.}
	\label{tab:notation_contribution}
\end{table}


\section{\boldmath The $B_d^0\to \pi^- \pi^+, B_s^0\to K^- K^+$ system} \label{sec:gammaDetermination}
We start with the $U$-spin system of the $B_d^0\to \pi^- \pi^+$ and $B_s^0\to K^- K^+$ decays. In the SM, these modes originate from tree ($T$), QCD penguin ($P$), exchange ($E$) and/or penguin-annihilation ($PA$) topologies, as listed in Table~\ref{tab:notation_contribution}. Interestingly, the CP asymmetries in this system can be used to extract the UT angle $\gamma$ using the $U$-spin symmetry \cite{Fleischer:2016ofb}, thereby giving a complementary determination to the usual extraction from pure tree decays, like $B\to D K$.  
We parameterize the decay amplitudes as \cite{Fleischer:2016ofb}  
\begin{align}
    A(B_s^0\to K^+ K^-) &= \sqrt{\epsilon}e^{i\gamma}\mathcal{C}^\prime \left[1 + \frac{1}{\epsilon}d' e^{i\theta'} e^{-i\gamma} \right]\ , \\
    A(B_d^0\to \pi^+\pi^-) &= e^{i\gamma}\mathcal{C}(1 + d e^{i\theta} e^{-i\gamma}) \ ,
\end{align}
where the primes indicate that we are dealing with a $\bar{b}\to \bar{s}$ transition. We define
\begin{equation}\label{eq:Cintro}
    \mathcal{C} \equiv \lambda^3 A R_b \left[T+E + P^{(ut)} + PA^{(ut)}\right] \ 
\end{equation}
with $P^{(qt)} \equiv P^{(q)} - P^{(t)}$, and in analogy for the $PA^{(qt)}$ contribution. We note that exchange and penguin-annihilation topologies contribute to both decays. There is a one-to-one correspondence of all decay topologies in this system that are related through the $U$-spin symmetry. Here $\gamma$ is the corresponding angle of the UT, $\lambda$ and $A$ are the Wolfenstein parameters of the CKM matrix \cite{Wolfenstein:1983yz,Buras:1994ec}, $\epsilon \equiv \lambda^2/(1-\lambda^2)$ and $R_b$ is one side of the UT. The hadronic parameters are given by  
\begin{equation} \label{eq:dthetaDef}
    d e^{i\theta}\equiv \frac{1}{R_b}\left[\frac{P^{(ct)} + PA^{(ct)}}{T+E+P^{(ut)}+PA^{(ut)}}\right]\ ,
\end{equation}
where $\theta$ is a CP-conserving strong phase and a similar expression holds for $d' e^{i\theta'}$. In the $U$-spin limit, we have the relation $d e^{i\theta}=d'e^{i\theta'}$ \cite{Fleischer:1999pa}. 
To quantify $U$-spin-breaking corrections, we introduce 
\begin{equation}\label{eq:xiDelta}
	\xi \equiv \frac{d'}{d}, \quad \Delta \equiv \theta' - \theta,
\end{equation}
where we have $\xi=1$ and $\Delta=0$ in the case of exact $U$-spin symmetry.

\subsection{Observables and inputs}\label{sec:obsin}
For neutral $B_q^0$ decays ($q=d,s$), CP violation is probed by the time-dependent rate asymmetry, caused by quantum-mechanical oscillations between the $B_q^0$ and $\bar{B}_q^0$ mesons \cite{Fleischer:2002ys}: 
\begin{align}\label{eq:acptime}
    \mathcal{A}_{\rm CP}(t) &= \frac{|A(B^0_q(t) \to  f)|^2 - |A(\bar{B}^0_q(t) \to f)|^2}{|A(B^0_q(t) \to  f)|^2 + |A(\bar{B}^0_q(t) \to f)|^2} \nonumber   \\
    &= 	\frac{\mathcal{A}_{\rm CP}^{\rm dir}(B_q\to f) \cos(\Delta M_q t) + 	\mathcal{A}_{\rm CP}^{\rm mix} \sin(\Delta M_q t)}{ \cosh(\Delta \Gamma_q t/2) + 	\mathcal{A}_{\rm CP}^{\Delta \Gamma} \sinh(\Delta \Gamma_q t/2)} \ ,
    \end{align}
where the mass and decay width differences between the “heavy” and “light” mass eigenstates are $\Delta M_q \equiv M^{(q)}_H - M^{(q)}_L$ and $\Delta \Gamma_q \equiv \Gamma^{(q)}_L - \Gamma^{(q)}_H$, respectively. We follow the definitions in \cite{Fleischer:2016ofb}, and refer the reader to this paper for expressions of the CP asymmetries in terms of the hadronic parameters $d$ and $\theta$ and their primed counterparts. The mixing-induced CP asymmetry is a key observable in these decays, as we will discuss in the next parts. Specifically, this introduces the CP-violating $B_q^0$--$\bar{B}_q^0$ mixing phases $\phi_q$ into the analyses.

\begin{table}[t]
	\centering
	\begin{tabular}{l | r}
	$\mathcal{A}_{\rm CP}^{\rm dir}(B_d^0 \to \pi^- \pi^+)$ 		& $-0.320 \pm 0.038$ \\
	$\mathcal{A}_{\rm CP}^{\rm mix}(B_d^0 \to \pi^- \pi^+)$ 	& $0.672 \pm 0.034$ \\
	\hline
	$\mathcal{A}_{\rm CP}^{\rm dir}(B_s^0 \to K^- K^+)$ 		& $0.172 \pm 0.031$ \\
	$\mathcal{A}_{\rm CP}^{\rm mix}(B_s^0 \to K^- K^+)$ 		& $-0.139 \pm 0.032$ \\
	$\mathcal{A}_{\rm CP}^{\Delta\Gamma}(B_s^0 \to K^- K^+)$ 		& $-0.897 \pm 0.087$ \\
	\hline
	$\mathcal{A}_{\rm CP}^{\rm dir}(B_d^0 \to \pi^- K^+)$ 	& $0.0831 \pm 0.0034$ \\
	\hline
	$\mathcal{A}_{\rm CP}^{\rm dir}(B_s^0 \to K^- \pi^+)$ 	& $-0.225 \pm 0.012$
	\end{tabular}
	\caption{Combined Run I \cite{LHCb:2018pff} and Run II LHCb results for the CP asymmetries of $B_{(s)} \to h^- h^+$ modes taken from \cite{LHCb:2020byh}.}
	\label{tab:LHCbPreACPs}
\end{table}
For $\phi_s$, we use a recent determination from $B_s^0 \to J/\psi \phi$ decays, where doubly Cabibbo-suppressed penguin corrections are taken into account \cite{Barel:2020jvf, Barel:2022wfr}:
\begin{equation} \label{eq:phisExperiment}
   \phi_s= -0.074^{+0.025}_{-0.024} = -(4.2 \pm 1.4)^\circ \ .
\end{equation}
Here
\begin{equation}\label{eq:phisdef}
    \phi_s \equiv \phi_s^{\rm SM} + \phi_s^{\rm NP} \ ,
\end{equation}
where $\phi_s^{\rm NP}$ describes possible CP-violating NP effects. For the SM predicition of $\phi_s$, we follow the strategy outlined in \cite{Barel:2020jvf}, where $\phi_s$ is obtained from a fit using only $|V_{cb}|$ and $|V_{ub}|$, $|V_{us}|$ and $\gamma$ (see \cite{Barel:2020jvf, Barel:2022wfr} for a detailed discussion). Specifically, we quote a very recent determination which found \cite{DeBruyn:2022zhw}
\begin{equation}
    \phi_s^{\rm SM}|_{\rm incl} = -(2.30\pm 0.13)^\circ \ , \quad\quad \phi_s^{\rm SM}|_{\rm excl} = -(2.08\pm 0.10)^\circ \ , 
\end{equation}
using $|V_{us}|$ from $K\ell 3$ decays and the inclusive and exclusive $|V_{ub}|$ and $|V_{cb}|$ determinations, respectively. In addition, \cite{DeBruyn:2022zhw} defines a hybrid scenario using $|V_{cb}|$ from inclusive decays and $|V_{ub}|$ from exclusive decays, leading to 
\begin{equation}
 \phi_s^{\rm SM}|_{\rm hybrid} = -(1.93\pm 0.10)^\circ \ .  
\end{equation}
In Sec.~\ref{sec:phisDetermination}, we compare our results with these determinations.

For the $B_d^0$--$\bar{B}_d^0$ mixing phase $\phi_d$, we use \cite{Barel:2020jvf, Barel:2022wfr}:
\begin{equation}\label{eq:phid}
   \phi_d= (44.4^{+1.6}_{-1.5})^\circ \ ,
\end{equation}
which follows from CP violation in $B_d^0 \to J/\psi K_S$ and takes also penguin corrections into account.

Besides the CP asymmetries, the branching ratios of these decays give additional information. However, in this case the overall normalization $\mathcal{C}$ does not drop out. This makes the branching ratio less clean to extract information on the hadronic parameters. 

For the neutral $B_q^0$ mesons, the measured ``experimental'' branching ratio differs from the ``theoretical'' definition as follows \cite{DeBruyn:2012wj}:
\begin{equation}\label{eq:theorybr}
    \mathcal{B}(B_q \to f)_{\rm theo} = \left[ \frac{1-y_q^2}{1+\mathcal{A}_{\rm CP}^{\Delta\Gamma}(B_q\to f) y_q}\right] \mathcal{B}(B_q\to f)_{\rm exp} \ ,
    \end{equation}
where 
\begin{equation}
    y_q \equiv \frac{\Delta\Gamma_q}{2 \Gamma_q} \equiv \frac{\Gamma_{\rm L}^{(s)}-\Gamma_{\rm H}^{(s)}}{2\Gamma_q} \ .
\end{equation}
Especially for the $B_s$ system, where \cite{Amhis:2022mac}
\begin{equation}
    y_s = 0.064 \pm 0.004 \ ,
\end{equation}
it is important to take this effect into account.
Finally, the most recent experimental results for the CP asymmetries in the $B_{(s)}\to h^- h^+$ modes from the LHCb Collaboration  \cite{LHCb:2020byh} are given in Table~\ref{tab:LHCbPreACPs}. We will use these measurements as inputs for our analysis.

\subsection{\boldmath Determination of $\gamma$}
\begin{figure}[t]
	\centering 
 	\includegraphics[width=0.5\textwidth]{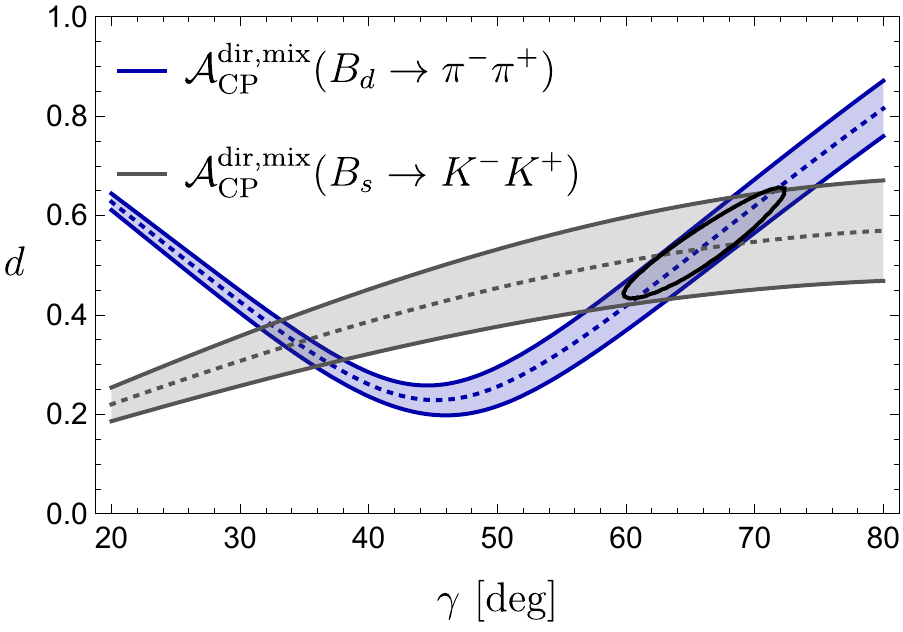}
	\caption{Determination of $\gamma$ from the CP asymmetries of the $B_d \to \pi^- \pi^+$ and $B_s \to K^- K^+$ modes from the LHCb data in Table~\ref{tab:LHCbPreACPs}.}
	\label{fig:originalplotcurrent}
\end{figure}

In the $B_s^0\to K^-K^+$ decay, penguin topologies play the dominated role, which may well be affected by new particles. The $B_d^0\to \pi^-\pi^+$ mode is governed by colour-allowed tree processes, due to the different CKM structure, but receives also significant penguin effects. It is therefore very interesting to compare the extraction of $\gamma$ from the pure tree-level $B\to D K$ decays with this $U$-spin system. 

Using the $B_q^0$--$\bar{B}_q^0$ mixing phases as an input, the direct and mixing-induced CP asymmetries can be used to express $d^{(')}$ as a function of $\gamma$, while eliminating $\theta^{(')}$ from the system. The current data in Table~\ref{tab:LHCbPreACPs} give the contours depicted in Fig.~\ref{fig:originalplotcurrent}.   
Assuming $U$-spin symmetry, the CP-violating observables from these decays can be combined to determine $\gamma$ \cite{Fleischer:1999pa, Fleischer:2016ofb} (see also \cite{LHCb:2013clb,LHCb:2014nbd}). Previously, the CP asymmetries alone did not allow for an extraction of $\gamma$, and the method had to supplemented with information from the branching ratios (see \cite{Fleischer:2016ofb}). Thanks to in particular the improved measurements of CP violation in the $B_s^0\to K^+K^-$ decay, we are now -- for the first time -- able to determine $\gamma$ using only the CP asymmetries. In total there are four solutions for $\gamma$, the intersections in Fig.~\ref{fig:originalplotcurrent} and two more that are shifted by $180^\circ$. The two solutions $\gamma = 34^\circ$ and $\gamma = -146^\circ$ have a value of $\left|\Delta\right| = \left| \theta' - \theta \right| = 86^\circ$, i.e., they would correspond to large $U$-spin-breaking effects, thereby disfavouring these two solutions. Of the remaining two solutions, $\gamma = -114^\circ$ is not favoured because it corresponds to values of $\theta$ and $\theta'$ close to $0^\circ$. However, based on the definition in \eqref{eq:dthetaDef} we expect an overall minus sign for $d^{(\prime)}$, i.e., a value of $\theta^{(\prime)}$ close to $180^\circ$, in view of the dominant contribution of the penguin topologies with internal top quarks.

First, we perform a $\chi^2$-fit assuming perfect $U$-spin symmetry, where we also include the observable $\mathcal{A}_{\rm CP}^{\Delta\Gamma}(B_s^0 \to K^-K^+)$. We find
\begin{equation}
    d = d' = 0.52_{-0.09}^{+0.13} \ 
    \end{equation}
and
\begin{equation}\label{eq:ourgam}
    \gamma = (65_{-5}^{+7})^\circ  \ .
\end{equation}
Allowing for $20\%$ $U$-spin breaking corrections in $d$ and $d'$ through $\xi=1\pm 0.2$ (see \cite{Fleischer:2016ofb}), we find
\begin{equation}\label{eq:gammauspin}
    \gamma_{U{\text{-spin}}} = (65^{+11}_{-7})^\circ \ .
\end{equation}
Comparing with \eqref{eq:ourgam}, we observe that the uncertainty has increased by a factor $1.5$. In the future, the $U$-spin corrections could be pinned down more precisely with the help of further experimental data, which would also reduce the uncertainty on $\gamma$.

\kkv{\cite{LHCb:2013clb,LHCb:2014nbd}  }

Our new determination in \eqref{eq:gammauspin} can be compared with other $\gamma$ determinations. For the pure tree-level determination from $B\to D K$ decays, we have \cite{LHCb:2021dcr}
\begin{equation}\label{eq:gamma}
    \gamma_{B\to D K}= (64.9 \pm 4.5)^\circ \ ,
\end{equation}
where we only consider the time-integrated analyses, i.e. excluding $B_s^0$ modes. We find that our determination is in impressive agreement with \eqref{eq:gamma}. The angle $\gamma$ can also be extracted from an isospin analysis of $B\to \pi\pi, \rho \pi, \rho\rho$ decays supplemented with $\phi_d$ in \eqref{eq:phid} as discussed in detail in \cite{DeBruyn:2022zhw}, finding
\begin{equation}\label{eq:isogam}
    \gamma_{\rm iso} = (72.6^{+4.3}_{-4.9})^\circ \ .
\end{equation}
These three determinations are consistent within the $1\,\sigma$ level, where we note that both our new determination in \eqref{eq:ourgam} and the one in \eqref{eq:isogam} could be affected by possible NP effects that enter via penguin topologies. In the remainder of this paper, we use $\gamma$ from \eqref{eq:gamma} as an input parameter.

Finally, we can also determine the CP-conserving strong phases
\begin{equation}\label{eq:thetagammadet}
    \theta = \left(147_{-10}^{+7}\right)^\circ \ , \quad \quad \theta'= \left(114_{-10}^{+9}\right)^\circ \ 
\end{equation}
and
\begin{equation}
    \Delta \equiv \theta'-\theta = \left(-33 ^{+11}_{-14}\right)^\circ\ ,
\end{equation}
where $\Delta$ vanishes in the $U$-spin limit. The obtained $U$-spin-breaking corrections are at the $20\%$ level. As these strong phases originate from non-factorizable processes, $U$-spin-breaking corrections at this level are not unexpected.

\subsection{Extracting the hadronic parameters} \label{subsec:ExtractingTheHadronicParameters}
Using $\gamma$ from \eqref{eq:gamma} allows us to determine the hadronic parameters $d, d'$ and their phases from the measurements of the mixing-induced and direct CP asymmetries of the $B_d^0\to \pi^+\pi^-, B_s^0\to K^+K^-$ system. We emphasize that this requires inputs for $\phi_d$ of $\phi_s$ in \eqref{eq:phid} and \eqref{eq:phisExperiment}. Following these lines, we find 
\begin{align}
	d &= 0.52 \pm 0.10, \quad \theta = (146.1 \pm 6.6)^\circ \label{eq:dthetaCurrent} \\
		d' &= 0.53 \pm 0.09, \quad \theta' = (113.6 \pm 12.0)^\circ \label{eq:dthetapCurrent}.
\end{align}
The values for $\theta$ and $\theta'$ are in excellent agreement with the determinations in \eqref{eq:thetagammadet}. Defining the $U$-spin-breaking parameters $\xi$ and $\Delta$ as in \eqref{eq:xiDelta}, we find 
\begin{equation}\label{eq:xi}
	\xi = 1.03 \pm 0.25, \quad \Delta = -(32.4 \pm 13.9)^\circ,
\end{equation}
These numbers allow for $U$-spin-breaking effects at the $20 \%$ level, strengthening the assumption used for the $\gamma$ determination above.


\section{\boldmath The $B_s^0\to K^- \pi^+, B_d^0\to \pi^- K^+$ system}
We continue with the $B_s^0\to K^- \pi^+$ and $B_d^0\to \pi^- K^+$ decays which are also related via the $U$-spin symmetry. In addition, these channels have flavour-specific final states. Consequently, there is no mixing-induced CP violation. The $B_d^0 \to \pi^- K^+$ and $B_s^0\to K^- K^+$ decays only differ by the flavour of the spectator quarks, with a similar situation for the $B_s^0 \to K^- \pi^+, B_d^0\to \pi^-\pi^+$ system.  Therefore, the significant difference between the direct CP asymmetries of these two $U$-spin systems already given in \eqref{eq:aCPdirDiff1} and \eqref{eq:aCPdirDiff2}, respectively, is rather striking. The SM explanation for this difference is given by sizeable exchange and penguin annihilation topologies present in the $B_d^0\to \pi^-\pi^+, B_s^0\to K^-K^+$ system but not in the $B_s^0\to K^- \pi^+, B_d^0 \to \pi^- K^+$ system. In the following, we present a new strategy to constrain these parameters through experimental data as they cannot be reliably calculated due to their non-pertubative nature.

\subsection{Hadronic parameters}
We parametrize the $B_s^0\to K^- \pi^+$ and $B_d^0\to \pi^- K^+$ decays as 
\begin{align} \label{eq:AmpBdtopiK}
    A(B_d^0 \to \pi^-K^+) &= \sqrt{\epsilon}e^{i\gamma} \tilde{\mathcal{C}}'\left[1+ \frac{1}{\epsilon} \tilde{d}'e^{i\tilde{\theta}'} e^{-i\gamma}\right] \  \\
 \label{eq:AmpBstopiK}
    A(B_s^0 \to K^-\pi^+)& = e^{i\gamma} \tilde{\mathcal{C}}\left[1- \tilde{d}e^{i{\tilde\theta}} e^{-i\gamma}\right] \ 
\end{align}
with
\begin{equation} \label{eq:defdtp}
    \tilde{d} e^{i\tilde{\theta}}\equiv \frac{1}{R_b} \left[\frac{\tilde{P}^{(ct)}}{\tilde{T} + \tilde{P}^{(ut)}} \right]  \ ;
\end{equation}
an analogous expression holds for $\tilde{d}^\prime e^{i\tilde{\theta}'}$. In the exact $U$-spin symmetry, these parameters are equal to each other. 

These flavour-specific decays exhibit only direct CP violation, which is defined in analogy with the time-dependent CP asymmetry \eqref{eq:acptime} as
\begin{equation}
    	\mathcal{A}_{\rm CP}^{\rm dir}(B^0_q \to f) =   \frac{|A(B^0_q \to  f)|^2 - |A(\bar{B}^0_q\to f)|^2}{|A(B^0_q \to  f)|f^2 + |A(\bar{B}^0_q \to f)|^2}  \ . 
\end{equation}
We can now use the $U$-spin symmetry to extract $\tilde{d}$ and $\tilde{\theta}$ from the direct CP asymmetries of the $B_d^0 \to \pi^- K^+$ and $B_s^0 \to K^- \pi^+$ decays. The current data give
\begin{equation} \label{eq:dthetatilUspin}
	\tilde{d} = 0.51 \pm 0.03, \quad \tilde{\theta} = (156.2 \pm 1.8)^\circ.
\end{equation}
There is a second solution, which is disfavoured based on arguments similar to those given in Sec.~\ref{sec:gammaDetermination}. In that section, we found that $U$-spin-breaking corrections could be as large as $20\%$ in the $B_s^0\to K^- K^+, B_d^0\to\pi^- \pi^+$ system. It is interesting to note that the $B_d^0\to \pi^- K^+, B_s^0\to K^- \pi^+$ system has only tree ($T$) and penguin ($P$) contributions. Due to the absence of non-factorizable exchange and penguin-annihilation topologies, we expect the $U$-spin symmetry to work even better for these decays.

Finally, we relate the hadronic paramters to those of the $B_d^0\to \pi^- K^+, B_s^0 \to K^- K^+$ system to parametrize the mismatch between the two $U$-spin systems:
\begin{equation} \label{eq:zetap}
	\tilde{d}' e^{i \tilde{\theta}'} = \zeta' d' e^{i \theta'}
\end{equation}
where
\begin{equation} \label{eq:zetapdef}
	\zeta' \equiv |\zeta'| e^{i \omega'} \equiv \frac{1+x'}{1+r_{PA}'} 
\end{equation}
with
\begin{equation} \label{eq:xrpadef}
	x^{(\prime)} \equiv |x^{(\prime)}| e^{i \sigma^{(\prime)}} \equiv \frac{E^{(\prime)} + PA^{(ut)(\prime)}}{T^{(\prime)} + P^{(ut)(\prime)}}, \quad r_{PA}^{(\prime)} \equiv |r_{PA}^{(\prime)}| e^{i \theta_{PA}^{(\prime)}} \equiv \frac{PA^{(ct)(\prime)}}{P^{(ct)(\prime)}}.
\end{equation}
The counterparts mediated by $b\to d$ transitions, the $B_s^0\to \pi^+ K^-$ and $B_d^0 \to \pi^- \pi^+$ decays, are related in a similar way:
\begin{equation} \label{eq:zeta}
	\tilde{d} e^{i\tilde{\theta}} = \zeta d e^{i \theta},
\end{equation}
where $\zeta$ is defined as $\zeta'$ but without the primes.

\subsection{Exchange and penguin annihilation contributions} \label{sec:EandPA}
The size of the exchange ($E$) and penguin-annihilation ($PA$) contributions denoted by the hadronic parameters $x^{(\prime)}$ and $r_{PA}^{(\prime)}$ and their phases can be determined through the new strategy illustrated in Fig.~\ref{fig:flowchart_EPA}:

\begin{figure}[t]
	\centering 
 	\includegraphics[width=\textwidth]{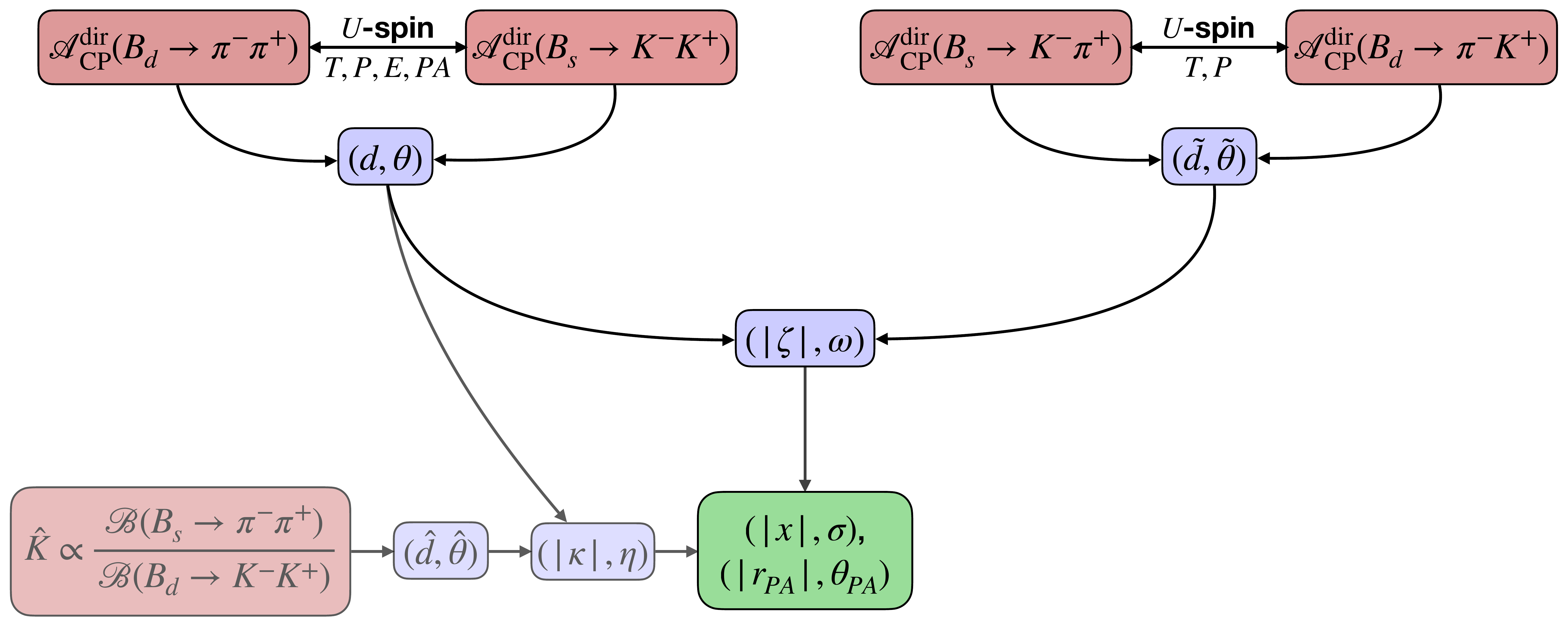}
	\caption{Our strategy to determine the impact of exchange ($E$) and penguin annihilation ($PA$) contributions parameterized by $x$, $r_{PA}$ and their CP-conserving complex phases. }
	\label{fig:flowchart_EPA}
\end{figure}
\vspace{0.2cm}
\noindent $\bullet$ {\bf{Step 1}}: We start by using the direct CP asymmetries only, which do not depend on the $B_q^0$--$\bar{B}_q^0$ mixing phases $\phi_{q}$. Assuming $U$-spin symmetry in the $B_d^0 \to \pi^- \pi^+, B_s^0 \to K^- K^+$ system, we can extract the hadronic parameters. Using in addition $\gamma$ from \eqref{eq:gamma}, we find 
\begin{equation}\label{eq:step1}
    d = d' =  0.39 \pm 0.05 , \qquad \theta = \theta' =  (140 \pm 5)^\circ  \ .
\end{equation}
It is interesting to compare these results for the hadronic parameters with those in~(\ref{eq:dthetaCurrent})~and~(\ref{eq:dthetapCurrent}) which were obtained using the mixing-induced CP asymmetries. For $d$, we observe a downward shift at the $1\, \sigma$ level, while the strong phase agrees with $\theta$ in~(\ref{eq:dthetaCurrent}).

\vspace{0.2cm}
\noindent $\bullet$ {\bf{Step 2}}:   
Using \eqref{eq:dthetatilUspin} and \eqref{eq:step1}, we determine $\zeta$:
\begin{equation}\label{eq:step2}
    |\zeta| \equiv \tilde{d}/d  = 1.3\pm 0.2\ , \quad \omega\equiv \tilde{\theta}-\theta =(16 \pm 5)^\circ \ . 
\end{equation}
These values provide a measure for the relative contribution of the $E^{(\prime)}$ and $PA^{(\prime)}$ topologies in the $B_d^0 \to \pi^- \pi^+$ ($B_s^0 \to K^- K^+$) decay. Since we used the direct CP asymmetries to obtain $d$ and $\tilde{d}$, this value of $\zeta$ automatically accommodates the differences in the direct CP asymmetries ~\eqref{eq:aCPdirDiff1}~and~\eqref{eq:aCPdirDiff2}. Their values indicate exchange and penguin annihilation effects at the level of $(20$--$30)\%$. The size of these effects is in the ballpark of general theoretical expectations (see e.g.~\cite{Gronau:1995hm}).

\vspace{0.2cm}
\noindent $\bullet$ {\bf{Step 3}}: Finally, we use $\zeta$ to obtain information on the size of the exchange and penguin annihilation contribution by determining $x$ and $r_{PA}$ and their strong phases using \eqref{eq:zetapdef}. This requires additional information. The $U$-spin related $B_s^0 \to \pi^- \pi^+, B_d^0 \to K^- K^+$ system, which receives only contributions from $E$ and $PA$ topologies, directly probes these parameters. This system therefore provides additional information to pinpoint the exchange and penguin annihilation effects. Let us, therefore, have a closer look at these decays.

\subsection{\boldmath The $B_s^0\to \pi^- \pi^+, B_d^0\to K^- K^+$ system}

The $U$-spin related $B_s^0 \to \pi^+\pi^-$ and $B_d^0\to K^+ K^-$ modes are parameterized as  \cite{Fleischer:2016ofb}
\begin{align}
  A(B_d^0\to K^+ K^-)  &= e^{i\gamma}\hat{\mathcal{C}} \left[1 - \hat{d} e^{i\hat\theta} e^{-i\gamma} \right] \\
A(B_s^0\to \pi^+\pi^-)   &= \sqrt{\epsilon} e^{i\gamma}\hat{\mathcal{C}}^\prime(1 + \frac{1}{\epsilon}\hat{d}^\prime e^{i\hat{\theta}'} e^{-i\gamma}) \ ,
\end{align}
with
\begin{equation} \label{eq:CDef}
    \hat{C}\equiv \lambda^3 A R_b \left[\hat{E} + \hat{PA}^{(ut)}\right] \ , \quad\quad \hat{d} e^{i\hat{\theta}}\equiv \frac{1}{R_b}\left[\frac{\hat{PA}^{(ct)}}{\hat{E}+\hat{PA}^{(ut)}}\right]\ ,
\end{equation}
and in analogy for $\hat{C}$ and $\hat{d}^\prime e^{i\hat{\theta}'}$. We define the ratio of their branching ratios as
\begin{align} \label{eq:KhatObservable}
    \hat{K} &\equiv \frac{1}{\epsilon}  \left[\frac{m_{B_s}}{m_{B_d}} \frac{\Phi(m_K/m_{B_d},m_K/m_{B_d})}{\Phi(m_\pi/m_{B_s},m_\pi/m_{B_s})} \frac{\tau_{B_d}}{\tau_{B_s}}\right] \frac{\mathcal{B}(B_s \to \pi^- \pi^+)_{\rm theo}}{\mathcal{B}(B_d \to K^- K^+)} \nonumber \\
&=\left|\frac{\hat{\mathcal{C}}^\prime}{\hat{\mathcal{C}}}\right|^2 \frac{1 + 2 (\hat{d}'/\epsilon) \cos\hat{\theta}' \cos\gamma + (\hat{d}'/\epsilon)^2}{1 - 2 \hat{d} \cos\hat{\theta} \cos\gamma + \hat{d}^2} \ ,
\end{align}
where 
\begin{equation} \label{eq:phaseSpaceFunction}
    \Phi(X,Y) = \sqrt{\left[1 - \left(X + Y\right)^2\right] \left[1 - \left(X - Y\right)^2\right]} 
\end{equation}
is the usual phase-space function. 
The theoretical branching ratio was already defined in \eqref{eq:theorybr}. Contrary to the CP asymmetries, this ratio depends on the overall normalization via the $\hat{C}'/\hat{C}$ factor. This quantity parametrizes the exchange and penguin-annihilation contributions, as seen in \eqref{eq:dthetaDef}, which are highly non-factorizable contributions. To estimate its size, we write \cite{Fleischer:2016ofb, Bobeth:2014rra}
\begin{equation}
    \frac{\hat{\mathcal{C}}}{\hat{\mathcal{C}}^\prime} \sim \frac{f_{B_d} f_K^2}{f_{B_s} f_\pi^2} \ , 
\end{equation}
where $f_{B_s}$ and $f_{B_d}$ are the $B_s$ and $B_d$ decays constants, and $f_K$ and $f_\pi$ those of kaons and pions, respectively.

\begin{table}
	\centering
	\begin{tabular}{l | r}
	$\mathcal{B}(B_d^0 \to \pi^- \pi^+)$ 		& $5.12 \pm 0.19$ \\
		$\mathcal{B}(B_s^0 \to K^- K^+)$ 		& $26.6 \pm 2.2$ \\
	\hline
	$\mathcal{B}(B_d^0 \to \pi^- K^+)$ 	& $19.6 \pm 0.5$ \\
	$\mathcal{B}(B_s^0 \to K^- \pi^+)$ 	& $5.8 \pm 0.7$ \\
	\hline
	$\mathcal{B}(B_d^0 \to K^- K^+)$ 		& $0.078 \pm 0.015$ \\
		$\mathcal{B}(B_s^0 \to \pi^- \pi^+)$ 		& $0.70 \pm 0.10$ \\
	\hline
	\end{tabular}
	\caption{Experimental branching ratios in $10^{-6}$ for the $B \to h^- h^+$ modes taken from \cite{Workman:2022ynf}.}
	\label{tab:Brsinput}
\end{table}

\begin{figure}
	\centering
	\subfloat{\includegraphics[height=0.3\textwidth]{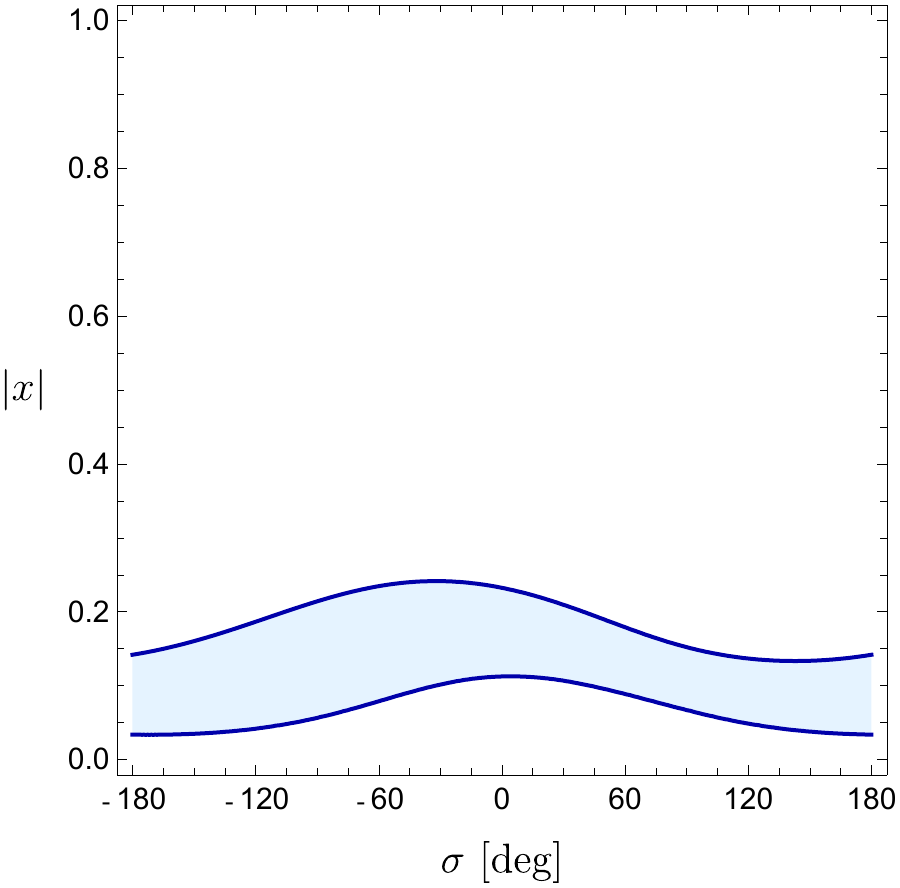}}
	\subfloat{\includegraphics[height=0.3\textwidth]{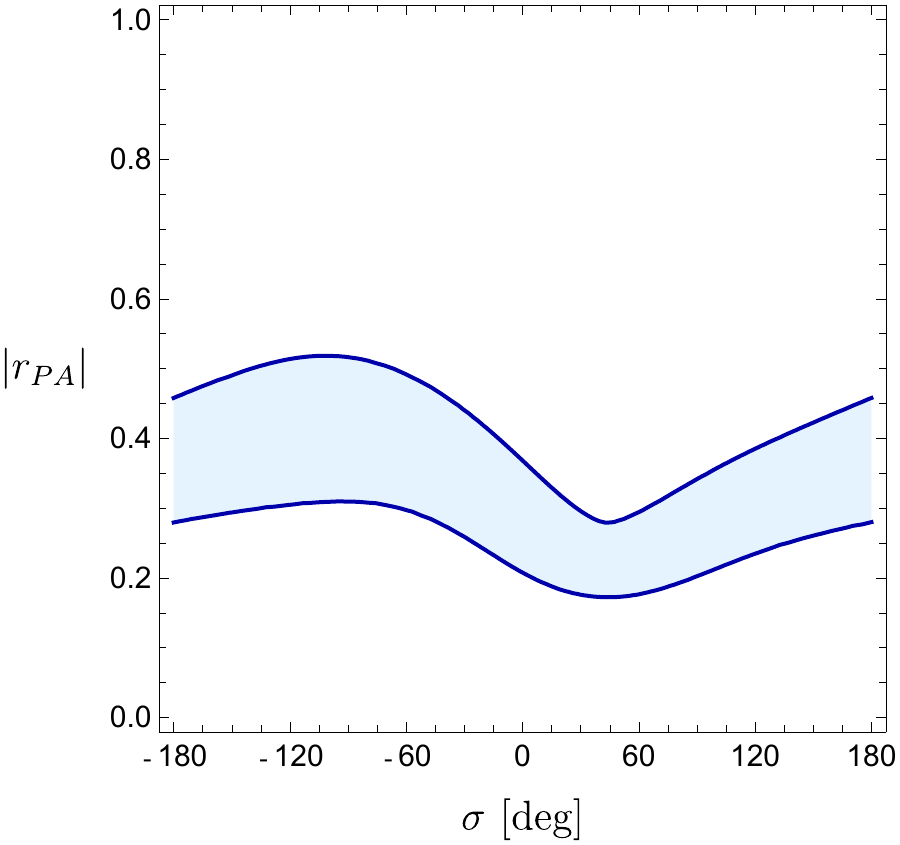}}
	\subfloat{\includegraphics[height=0.3\textwidth]{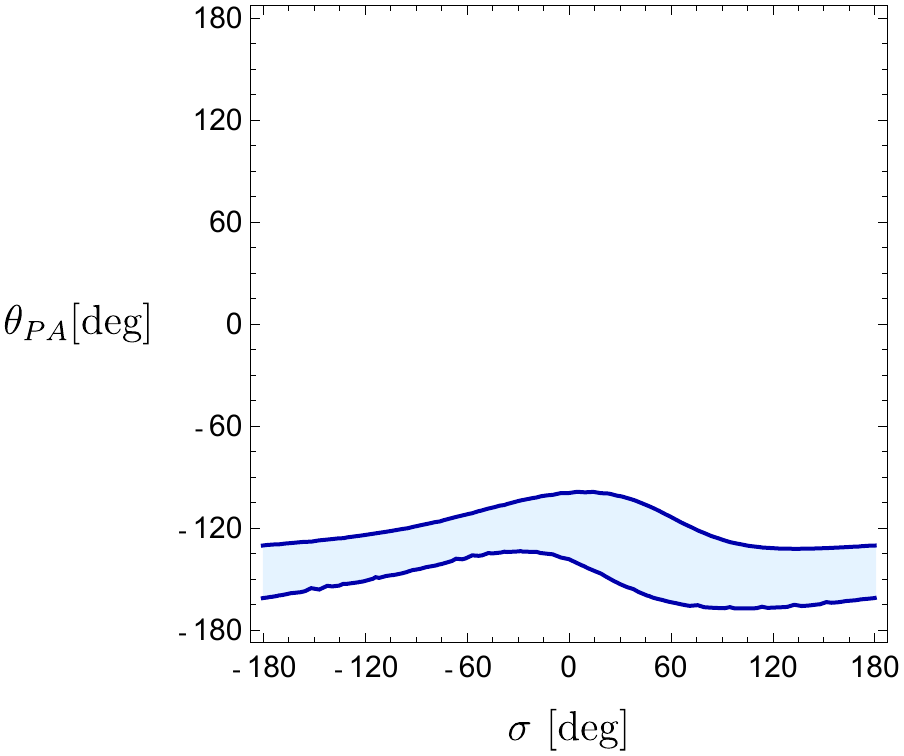}}
	\caption{The allowed $1\, \sigma$ regions from current data for different combinations of the exchange ($x, \sigma$) and penguin-annihilation ($r_{PA}, \theta_{PA}$) parameters.}
	\label{fig:EPAfitRegions}
\end{figure}

Utilizing the direct and mixing-induced CP asymmetries of this system, we could directly determine $x$ and $r_{PA}$ from the experimental data. Unfortunately, these asymmetries have not yet been measured, and therefore we will use the ratio of branching ratios as an alternative. In order to do so, it is convenient to use the relation between $\hat{d}$ and $\hat{\theta}$ and their $B_d^0 \to \pi^+\pi^-$ counter part:
\begin{equation}
    \hat{d} e^{i\hat{\theta}}= \kappa d e^{i\theta} \ ,
\end{equation}
where we defined
\begin{equation}\label{eq:kappadthetahat}
    \kappa \equiv |\kappa|e^{i\eta} = \left(\frac{r_{PA}}{1+r_{PA}}\right)\left( \frac{1+x}{x}\right) \ ,
\end{equation}
with the CP-conserving strong phase $\eta$. We can now express $\hat{K}$ in terms of $|\kappa|$ and $\eta$ with the help of the values for $d$ and $\theta$ obtained in \eqref{eq:step1}. Using the experimental values for the branching ratios given in Table~\ref{tab:Brsinput} and combining them with the obtained constraint on $\zeta$ in \eqref{eq:step2} gives three constraints on four parameters. In Fig.~\ref{fig:EPAfitRegions}, we show the picture for $|x|$, $|r_{PA}|$ and $\theta_{PA}$ as function of the phase $\sigma$ following from the current data. We find constraints around $20\%$ and $30\%$ for $x$ and $r_{PA}$, respectively, while the phase $\theta_{PA}$ lies within the interval $(90^\circ$--$180^\circ)$. 

\vspace{0.2cm}
\noindent In conclusion, the difference in the direct CP asymmetries in \eqref{eq:aCPdirDiff1} and \eqref{eq:aCPdirDiff2} can be accommodated by exchange and penguin annihilation effects at the level of $(20$--$30)\%$ of the overall amplitudes. In the future, with measurements of the CP asymmetries in the $B_s^0\to \pi^-\pi^+, B_d^0\to K^- K^+$ system, the size of the exchange and penguin annihilation parameters can be fully determined.

\section{\boldmath Extracting $\phi_s$ from the $B_s^0 \to K^- K^+$ decay}\label{sec:phisDetermination}
As an alternative to the determination of $\gamma$, the penguin-dominated $B_s^0 \to K^-K^+$ decay mode can also be used to determine the $B_s^0$--$\bar{B}_s^0$ mixing phase $\phi_s$. This avenue is particularly interesting because NP could influence this decay in a different way as the $B_s^0 \to J/\psi \phi$ channel, which is dominated by colour-suppressed tree topologies with a smallish penguin contribution, yielding the value for $\phi_s$ given in \eqref{eq:phisExperiment}.

The CP asymmetries of $B_s^0\to K^- K^+$ allow the extraction of the effective phase
\begin{equation}
    \sin\phi_s^{\rm eff} = \frac{ \mathcal{A}_{\rm CP}^{\rm mix}(B_s^0\to K^-K^+)}{ \sqrt{1-\left[\mathcal{A}_{\rm CP}^{\rm dir}(B_s^0\to K^-K^+)\right]^2}} \ ,
\end{equation}
which is defined as
\begin{equation}
    \phi_s^{\rm eff} \equiv \phi_s + \Delta \phi_{KK} \ ,
\end{equation}
where $\phi_s$ is the $B_s^0$--$\bar{B}_s^0$ mixing phase given in \eqref{eq:phisdef} and $\Delta\phi_{KK}$ a hadronic phase shift. 

The LHCb measurements of the CP asymmetries of $B_s^0 \to K^- K^+$  given in Table~\ref{tab:LHCbPreACPs} correspond to an effective mixing phase of 
\begin{equation}\label{eq:phiseff}
	\phi_s^{\rm eff} = -(8.1 \pm 1.9)^\circ,
\end{equation}
which has an impressively small uncertainty. The final step to extract $\phi_s$ is to determine $\Delta\phi_{KK}$, which depends on the hadronic parameters $d'$ and $\theta'$ as follows:
\begin{equation}\label{eq:deltaphikk}
    \tan\Delta \phi_{KK} = 2 \epsilon \sin\gamma \left[\frac{d' \cos\theta'+ \epsilon \cos\gamma}{d^{\prime 2}+ 2 \epsilon d' \cos\theta' \cos\gamma + \epsilon^2 \cos 2\gamma}\right] \ ,
\end{equation}
where $\epsilon$ was introduced after \eqref{eq:Cintro}.

In~\cite{Fleischer:2016ofb}, we presented a strategy using semileptonic decays to determine $\Delta\phi_{KK}$ in an optimal way, which we disscuss in the following.  

\subsection{\boldmath Strategy using semileptonic $B_{(s)}$ decays}
One of the key aspects of this new strategy is the use of (double) ratios of non-leptonic and semileptonic $B_{(s)}$ decay rates \cite{Fleischer:2016ofb}: 
\begin{equation}\label{eq:defRpi}
	R_{\pi}  \equiv  \frac{ \Gamma(B_d^0\to \pi^-\pi^+)}{|d\Gamma(B^0_d\rightarrow \pi^- \ell^+ 
	\nu_\ell)/dq^2|_{q^2=m_\pi^2}} \ , 	\quad\quad R_{K}  \equiv  \frac{ \Gamma(B_s^0\to K^- K^+)_{\rm theo}}{|d\Gamma(B^0_s\rightarrow K^- \ell^+ 
	\nu_\ell)/dq^2|_{q^2=m_K^2}} \ .
\end{equation}
Expressing the decay rates in terms of the hadronic parameters gives 
\begin{equation}\label{eq:defRpi2}
	R_{\pi} = 6\pi^2 |V_{ud}|^2 f_\pi^2 X_{\pi} r_\pi |a_{\rm NF}|^2 \ , \quad\quad	R_K = 6\pi^2 |V_{us}|^2 f_K^2 X_{K} r_K |a_{\rm NF}^\prime|^2 \ , 
\end{equation}
where $f_{\pi, K}$ are the meson decay constants, and the $X_{\pi,K}$ are ratios of phase-space and form factors defined as
\begin{equation}\label{eq:Xdef}
    X_\pi \equiv \frac{(m_{B_d}^2 - m_\pi^2)^2}{m_{B_d}^2 ( m_{B_d}^2-4m_\pi^2)}  \left[ \frac{F_0^{B_d\pi}(m_\pi^2)}{F_1^{B_d\pi}(m_\pi^2)} \right]^2 \ ;
\end{equation}
a similar expression with straightforward replacements holds for $X_K$. The form factors $F_0$ and $F_1$ are defined in \cite{Fleischer:2016ofb}. At $q^2 = 0$ their ratio is exactly one due to kinematic constraints. Since for $q^2=m_\pi^2$ and $q^2= m_K^2$ we are close to this situation, the form factor dependence essentially drops out in $X_\pi$ and $X_K$.

The hadronic parameters enter via
\begin{equation}\label{eq:defrpi}
r_\pi \equiv 1+d^2-2d\cos\theta\cos\gamma		 \ .
\end{equation}
Equivalently for the $B_s$ decay defined with a prime, the hadronic parameters enter through
\begin{equation}\label{eq:defrk}
    r_K \equiv 1+\left(\frac{d^{\prime}}{\epsilon}\right)^2+2 \frac{d^\prime}{\epsilon} \cos\theta^\prime \cos\gamma .
\end{equation}
Here $a_{\rm NF}$ parametrises the non-factorisable contributions through
\begin{equation} \label{eq:defaNFd}
	a_{\rm NF} \equiv (1+r_P)(1+x) a^T_{\rm NF} \ ,
\end{equation}
where $r_P = P^{(ut)}/T$ denotes the ratio of penguin to tree topologies. Finally, $a_{\rm NF}^T$ describes the non-factorisable contributions to the colour-allowed tree topology, which can be computed in QCD factorisation \cite{Beneke:1999br,Beneke:2001ev}.

Using $\phi_d$ and $\gamma$ as input parameters allows us to extract the penguin parameters $d$ and $\theta$ from the CP asymmetries of $B_d^0 \to \pi^- \pi^+$, thereby allowing us to determine $r_\pi$ in a theoretically clean way. These values were actually determined already in Eq.~(\ref{eq:dthetaCurrent}). Taking then the ratio of $R_\pi$ and ${R}_K$ yields
\begin{equation} \label{eq:rkTh}
    {r}_K = \frac{{R}_K}{R_\pi} \left( \frac{|V_{ud}| f_\pi}{|V_{us}| f_K} \right)^2 \frac{X_\pi}{{X}_K} \left( {\xi}_\text{NF}^a \right)^2 r_\pi \ ,
\end{equation}
leaving us with a function of the hadronic parameters $d'$ and $\theta'$ as given in Eq.~(\ref{eq:defrk}). Because the ratio of the CKM factors and decay constants has been determined very precisely \cite{Workman:2022ynf}, the only remaining theoretical uncertainty enters through
\begin{equation}\label{eq:xidef}
    {\xi}_\text{NF}^a \equiv \left| \frac{1 + r_P}{1 + {r}'_P} \right| \left| \frac{1 + x}{1+x'} \right| \left| \frac{a_{\rm NF}^T}{{a}_{\rm NF}^{T'}} \right|,
\end{equation}
which parametrises the non-factorisable $U$-spin-breaking contributions. In the exact $U$-spin symmetry, $\xi_{\rm NF}^a=1$. Thanks to the use of semileptonic ratios, the non-factorisable effects only enter in the form of double ratios. This leaves a very favourable structure from the perspective of potential $U$-spin-breaking corrections, because these effects do not enter linearly. Thanks to our new analysis, we can re-evaluate the estimate of the uncertainty on $\xi_{\rm NF}^a$ we already made in \cite{Fleischer:2016ofb}. To this extend, we write as in \cite{Fleischer:2016ofb}:
\begin{equation}
    \Xi_x \equiv \left|\frac{1+x}{1+x'}\right| = 1 + x \xi_x + \mathcal{O}(x^2) \ ,
\end{equation}
where $\Xi_x$ measures the $U$-spin-breaking corrections in the exchange topologies. Including $U$-spin-breaking corrections at the level of $20\%$ yields with a conservative estimate of $x \sim 0.1\pm 0.1$ from Fig.~\ref{fig:EPAfitRegions}, we find a $4\%$ uncertainty from $\Xi_x$. To estimate the remaining uncertainty on $\xi_{\rm NF}^a$, we follow our previous analysis \cite{Fleischer:2016ofb} in which we found a $1\%$ uncertainty from $a_{\rm NF}^T$ and a $5\%$ uncertainty from the $r_P$ part. The latter requires also information from the charged $B^+$ modes and we explicitly checked that the estimates we made in \cite{Fleischer:2016ofb} are in agreement with current data. Adding these uncertainty contributions in quadrature, we find 
\begin{equation}\label{eq:xiin}
    \xi_{\rm NF}^a = 1.00 \pm 0.07 \ .
\end{equation}

Finally, having obtained $r_K$ using the experimental $R_K/R_\pi$ ratios, we have fixed the function of $d'$ and $\theta'$ given in Eq.~(\ref{eq:defrk}). Another function of $d'$ and $\theta'$ is provided by the direct CP asymmetry of $B_s^0 \to K^-K^+$. Consequently, we have sufficient information to extract $d'$ and $\theta'$. Using $\Delta\phi_{KK}$ in \eqref{eq:deltaphikk}, it is then possible to calculate the hadronic phase shift and convert the effective mixing phase $\phi_s^{\rm eff}$ from \eqref{eq:phiseff} into the $B_s^0$--$\bar{B}_s^0$ mixing phase $\phi_s$.  

The semileptonic $B_s^0 \to K^- \ell^+ \nu_\ell$ decay has recently been observed for the first time by the LHCb collaboration \cite{LHCb:2020ist}. Measured is the integrated rate in different regions of $q^2$, but unfortunately results for the differential rate at $q^2=m_K^2$ have not yet been reported. Therefore $R_K$ and consequently $\Delta \phi_{KK}$ cannot yet be determined with this strategy. Once available, this method would be the most favourable to pursue because any experimental improvement directly leads to a more precise determination of $\phi_s$. In \cite{Fleischer:2016ofb}, we discussed in detail how the theoretical uncertainty, i.e. the uncertainty on $\xi_{\rm NF}^a$, compares to the experimental uncertainties. Taking the $\xi_{\rm NP}^a$ in \eqref{eq:xiin} gives a theoretical uncertainty of only $0.8^\circ$ on $\Delta\phi_{KK}$.

\subsection{Alternative strategy}
Since the differential semileptonic $B_s^0\to K^- \ell^+ \nu_\ell$ rate is not yet available, it is interesting to replace this quantity by the $B_d^0\to \pi^- \ell^+ \nu_\ell$ rate. In \cite{Fleischer:2016ofb}, we used the $B_d^0 \to \pi^- K^+$ channel instead of $B_s^0 \to K^- K^+$ with the semileptonic $B_d^0\to \pi^-\ell^+ \nu_\ell$ differential rate. The downside of this approach is that the $B_d^0 \to \pi^- K^+$ mode only receives tree and penguin contributions and no penguin-annihilation and exchange contributions. Consequently, there is a mismatch with the $B_d^0 \to \pi^-\pi^+$ decay, resulting in a modified expression for $\xi_{\rm NP}^a$. This strategy was discussed in detail in \cite{Fleischer:2016ofb}, but there we neglected exchange and penguin annihilation contributions for simplicity. In view of the new results obtained in Sec.~\ref{sec:EandPA}, these effects should be taken into account. However, the current constraints on these contributions are such that a conservative analysis of these effects renders large uncertainties.

We therefore focus on a second alternative using only the nonleptonic $B_s^0\to K^-K^+$ and $B_d^0\to \pi^-\pi^+$ rates. We define 
\begin{equation} \label{eq:tilrkTh2}
    {r}_K =     \left|\frac{\mathcal{C}}{\mathcal{C}'}\right|^2 K r_\pi \ ,
\end{equation}
where $K$ is the ratio of branching ratios defined as \footnote{The definition in \eqref{eq:kObservable} differs from that in \cite{Fleischer:2016ofb} because we have kept the factor  $\mathcal{C}/\mathcal{C}'$ on the right-hand side for convenience.}
\begin{align} \label{eq:kObservable}
    K &\equiv \frac{1}{\epsilon} \left[\frac{m_{B_s}}{m_{B_d}} \frac{\Phi(m_\pi/m_{B_d},m_\pi/m_{B_d})}{\Phi(m_K/m_{B_s},m_K/m_{B_s})} \frac{\tau_{B_d}}{\tau_{B_s}}\right] \frac{\mathcal{B}(B_s^0 \to K^- K^+)_{\rm theo}}{\mathcal{B}(B_d^0 \to \pi^-\pi^+)} 
    \nonumber \\
&=  \left|\frac{\mathcal{C}^\prime}{\mathcal{C}}\right|^2\frac{1 + 2(d'/\epsilon) \cos\theta' \cos\gamma + (d'/\epsilon)^2}{1 - 2 d \cos\theta \cos\gamma + d^2} ,
\end{align}
and $\Phi$ is given in \eqref{eq:phaseSpaceFunction}. The difference with respect to the semileptonic strategy is the appearance of the ratio $\mathcal{C}/\mathcal{C}^\prime$, which introduces a dependence on the form factors. We have
\begin{equation}\label{eq:cfact}
    \left|\frac{\mathcal{C}}{\mathcal{C}'}\right| = \frac{f_\pi}{f_K} \left[\frac{m_{B_d}^2 - m_\pi^2}{m_{B_s}^2 - m_K^2}\right] \left[\frac{F_0^{B_d \pi}(m_\pi^2)}{F_0^{B_s K}(m_K^2)}\right] \xi^a_{\rm NF}\ 
\end{equation}
with $f_K/f_\pi = 1.1928\pm 0.0026$ \cite{Rosner:2015wva}. For the form factor ratio we use the light-cone sum rule calculation in \cite{Khodjamirian:2017fxg} finding
\begin{equation}\label{eq:ffun}
\frac{F_0^{B_s K}(m_K^2)}{F_0^{B_d\pi}(m_\pi^2)} = 1.12 \pm 0.11 \ .
\end{equation}
This result has a rather sizeable uncertainty, in view of having only spectator-quark effects, and is the largest source of the theoretical uncertainty for this alternative strategy. We stress that when using the semileptonic rates, the form factors only enter via the ratio $X_{\pi, K}$ ratios defined in \eqref{eq:Xdef}. As discussed there, this makes the semileptonic strategy, which is essentially free of form-factor uncertainties. Therefore, it is much cleaner than the here discussed alternative.  

\begin{figure}[t]
	\centering 
 	\includegraphics[width=0.5\textwidth]{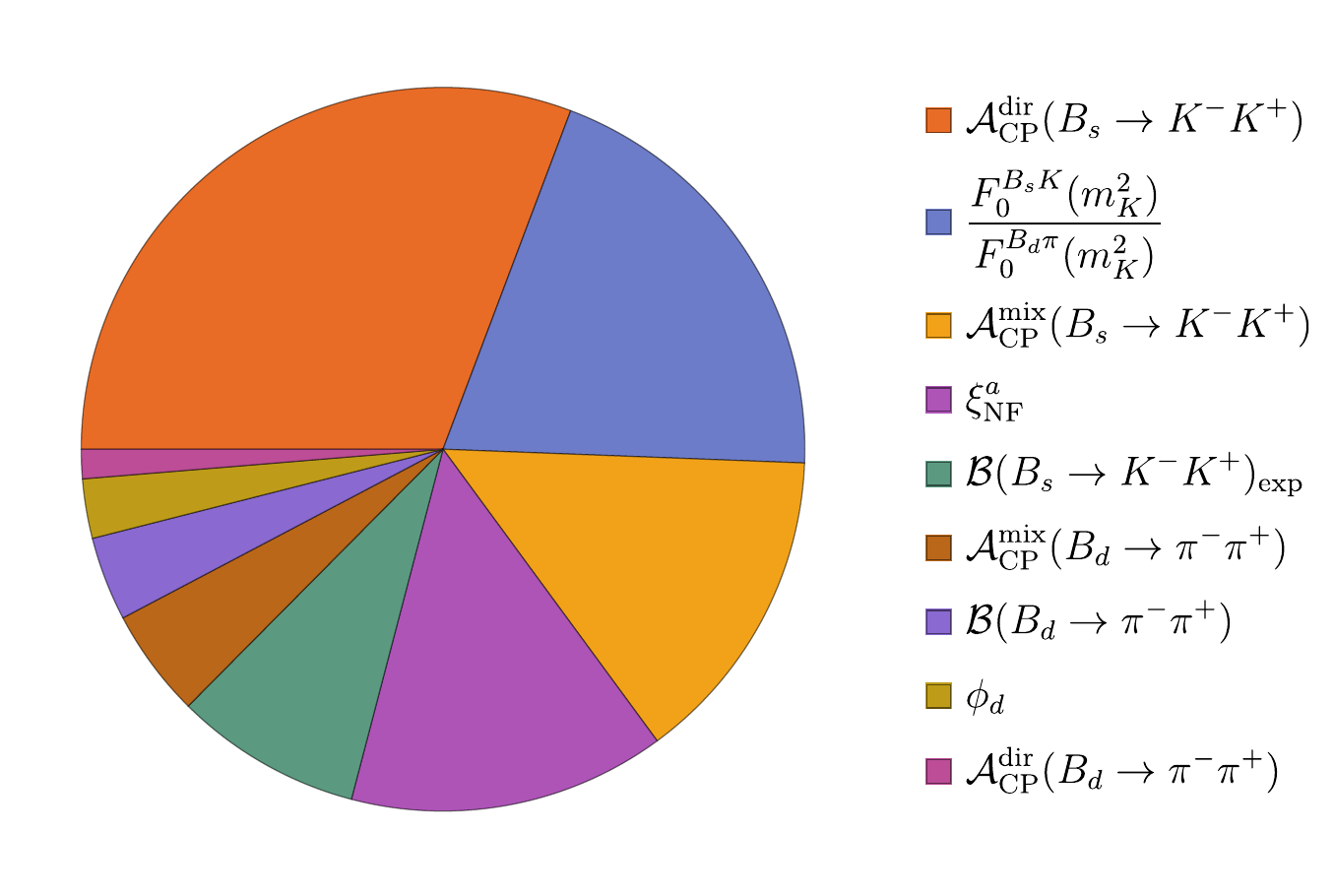}
	\caption{Relative contributions to the uncertainty of the value of $\phi_s$ in Eq.~(\ref{eq:phisfromK}). }
	\label{fig:errbudgetphisK}
\end{figure}

Using the experimental values for the branching ratios in Table~\ref{tab:Brsinput}, we find
\begin{equation}
    K|_{\rm exp} =  105.3 \pm 9.6 \ .
\end{equation}
Combining this with $(d, \theta)$ from the $B_d^0\to \pi^- \pi^+$ decay in \eqref{eq:dthetaCurrent}, $\gamma$ from \eqref{eq:gamma} and $\xi_{\rm NF}^a$ from \eqref{eq:xiin}, we find 
\begin{equation}
    d' = 0.52 \pm 0.06\  , \quad\quad \theta' = (116.5 \pm 28.0)^\circ \ ,
\end{equation}
which, using Eq.~\eqref{eq:deltaphikk}, gives
\begin{equation}
    \Delta\phi_{KK} = -(4.5 \pm 5.3)^\circ \ .
\end{equation}
Finally, using the measured value of $\phi_s^{\rm eff}$ in \eqref{eq:phiseff}, we obtain
\begin{equation} \label{eq:phisfromK}
    \phi_s = -(3.6 \pm 5.7)^\circ \ . 
\end{equation}
The relative contributions to this uncertainty are given in Fig.~\ref{fig:errbudgetphisK}. We observe that our new $\phi_s$ determination is dominated by the experimental uncertainties on the $B_s^0\to K^-K^+$ CP asymmetries, followed by the theoretical uncertainty on the form factors. Consequently, this analysis could this still be improved with updated experimental measurements.

Our new determination of $\phi_s$ in \eqref{eq:phisfromK} should be compared with the experimental determination from $B_s^0\to J/\psi \phi$ in \eqref{eq:phisExperiment}. We find a remarkable agreement between the two determinations. In addition, our new determination also agrees with the SM predictions given in Sec.~\ref{sec:obsin} but still leaves significant room for new physics. 

\section{Conclusions} \label{sec:Conclusion}
The first observation of CP violation in $B_s^0\to K^-K^+$ by the LHCb collaboration allows the determination of the UT angle $\gamma$ and the $B_s^0$--$\bar{B}^0_s$ mixing phase $\phi_s$ from this penguin-dominated decay. Using its $U$-spin partner mode $B_d^0\to \pi^-\pi^+$, we extract $\gamma = (65^{+11}_{-7})^\circ$ for the first time from only the CP asymmetries. This new determination is in excellent agreement with the results from pure tree-level $B\to D K$ decays.

We pointed out a surprising difference in the direct CP asymmetries between the $B_s^0 \to K^- K^+, B_d^0 \to \pi^- K^+$ and the $B_d^0 \to \pi^- \pi^+, B_s^0 \to K^- \pi^+$ decays, respectively, given in \eqref{eq:aCPdirDiff1} and \eqref{eq:aCPdirDiff2}. Using a new strategy, we found that this pattern can be accommodated through exchange and penguin-annihilation topologies at the level of $20\%$. Consequently, our results do not indicate any anomalous enhancement of these topologies through large non-factorizable effects. Future measurements of the CP asymmetries of the $U$-spin-related $B_d^0\to K^-K^+, B_s^0\to \pi^-\pi^+$ decays would allow a direct determination of these contributions from data.

Our new analysis shows that exchange and penguin annihilation contributions have to be included when analyzing the current data. We demonstrated that $\phi_s$ can be obtained using semileptonic $B_s^0$ and $B_d^0$ differential rates \cite{Fleischer:2016ofb}. In this way, the exchange and penguin-annihilation terms only enter through ratios and would cancel in the $U$-spin limit. Because measurements of the required differential rate of $B_s^0\to K^- \ell^+ \nu_\ell$ are currently not available, we presented a new strategy using a theoretical calculation of the required $SU(3)$-breaking form-factor ratio  instead. Employing the currently available data, we found $\phi_s = -(3.6 \pm 5.7)^\circ$, which is in full agreement with the determination of this phase
from CP violation in $B_s^0\to J/\psi \phi$ decays.

It will be exciting to see how the determination of $\phi_s$ from $B_s^0\to K^-K^+$ will develop in the future when its potential can be fully exploited by utilising
also the $B_s^0\to K^-\ell^+\nu_\ell$ differential rate. We would like to stress again that this strategy remains the cleanest way to extract $\phi_s$ from the penguin-dominated $B_s^0 \to K^- K^+$ decay. It is therefore important to measure the semileptonic rate at the required kinematic point.

Moving towards highest precision, the key question is whether the corresponding result for $\phi_s$ will develop a discrepancy with the value following from $B_s^0\to J/\psi \phi$ analyses, taking also smallish penguin effects into account. Such a phenomenon would arise in the case of CP-violating new-physics contributions to the amplitudes of these decays, where the penguin-dominated $B_s^0\to K^-K^+$ channel appears particularly sensitive.

Should the results continue to be in perfect agreement with each other, the key target would be the comparison with the SM predictions of $\phi_s$. In case of an established discrepancy, we would require new CP-violating contributions to $B_s^0$--$\bar{B}^0_s$ mixing. In specific scenarios for physics beyond the SM, both kinds of new physics effects may well arise. The corresponding studies at the future high-precision frontier have the exciting potential to finally establish new sources of CP violation and should move into the experimental spotlight.


\section*{Acknowledgements}
This research has been supported by the Netherlands Organisation for Scientific Research
(NWO). 

\bibliographystyle{JHEP} 
\bibliography{refs.bib}
\end{document}